\documentclass[twocolumn]{revtex4}
\usepackage{graphicx}

\def\duzomniejsze{<\kern-.7mm<}
\def\duzowieksze{>\kern-.7mm>}

\def\textbf#1{{\bf #1}}
\def\beq{\begin{equation}}
\def\eeq{\end{equation}}
\def\be{\begin{equation}}
\def\ee{\end{equation}}
\def\ben{\begin{eqnarray}}
\def\een{\end{eqnarray}}
\def\beqa{\begin{eqnarray}}
\def\eeqa{\end{eqnarray}}
\def\eea{\end{array}}
\def\bea{\begin{array}}
\newcommand{\bei}{\begin{itemize}}
\newcommand{\eei}{\end{itemize}}
\newcommand{\bee}{\begin{enumerate}}
\newcommand{\eee}{\end{enumerate}}

\def\>{\rangle}
\def\<{\langle}

\begin{document}

\title{Robust and fragile Werner states in the collective dephasing}

\begin{abstract}
We investigate the concurrence and Bell violation of the standard
Werner state or Werner-like states in the presence of collective
dephasing. It is shown that the standard Werner state and certain
kinds of Werner-like states are robust against the collective
dephasing, and some kinds of Werner-like states is fragile and
becomes completely disentangled in a finite-time. The threshold
time of complete disentanglement of the fragile Werner-like states
is given. The influence of external driving field on the
finite-time disentanglement of the standard Werner state or
Werner-like states is discussed. Furthermore, we present a simple
method to control the stationary state entanglement and Bell
violation of two qubits. Finally, we show that the theoretical
calculations of fidelity based on the initial Werner state
assumption well agree with previous experimental results.

PACS numbers: 03.65.Ud, 03.67.-a, 05.40.Ca
\end{abstract}
\author{Shang-Bin Li}\email{Stephenli74@yahoo.com.cn}, \author{Jing-Bo Xu}

\affiliation{Chinese Center of Advanced Science and Technology
(World Laboratory), P.O.Box 8730, Beijing, People's Republic of
China;} \affiliation{Zhejiang Institute of Modern Physics and
Department of Physics, Zhejiang University, Hangzhou 310027,
People's Republic of China}

\maketitle


Quantum entanglement plays a crucial role in quantum information
processes \cite{Nielsen2000}. Entanglement can exhibit the nature
of a nonlocal correlation between quantum systems that have no
classical interpretation. However, real quantum systems will
unavoidably be influenced by surrounding environments. The
interaction between the environment and quantum systems of
interest can lead to decoherence. Certain kind of the decoherence
is the collective dephasing, which occurs in the physical systems
such as trapped ions, quantum dots, or atoms inside a cavity.
Collective dephasing allows the existence of the so-called
decoherence-free subspace \cite{Zanardi2001}.

Recently, the Werner or Werner-like states
\cite{Werner1989,Munro2001,Ghosh2001,Wei2003} has intrigued many
interests for the applications in quantum information processes.
Lee and Kim have discussed the entanglement teleportation via the
Werner states \cite{Lee2000}. Hiroshima and Ishizaka have studied
the entanglement of the so-called Werner derivative, which is the
state transformed by nonlocal unitary-local or nonlocal-operations
from a Werner state \cite{Hiroshima2000}. Miranowicz has examined
the Bell violation and entanglement of Werner states of two qubits
in independent decay channels \cite{Miranowicz2004}. The
experimental preparation and characterization of the Werner states
have also been reported. An experiment for preparing the Werner
state via spontaneous parametric down-conversion has been put
forward \cite{Zhang2002}. Altepeter \textit{et al.} have
experimentally produced the Werner state and applied it in the
ancilla-assisted process tomography \cite{Altepeter2003}. Barbieri
\textit{et al.} have presented a novel technique for generating
and characterizing two-photon polarization Werner states
\cite{Barbieri2004}, which is based on the peculiar spatial
characteristics of a high brilliance source of entangled pairs.

The disentanglement of entangled states of qubits is a very
important issue for quantum information processes, such as the
solid state quantum computation. Yu and Eberly have found that the
time for decay of the qubit entanglement can be significantly
shorter than the time for local dephasing of the individual qubits
\cite{Yu2002,Yu2003}. In this paper, we investigate the
entanglement and Bell violation of the standard Werner state or
Werner-like states in the presence of collective dephasing. The
entanglement quantified by the concurrence and Bell violation of
the collective dephasing Werner-like state are analyzed. We find
that the standard Werner state and certain kinds of Werner-like
states are robust against the collective dephasing, and some kinds
of Werner-like states are fragile and become completely
disentangled in a finite-time. The threshold time for the complete
disappearance of the entanglement of the fragile Werner-like
states is obtained. We also provide an explicit example to clarify
how the pure maximally entangled states of two qubits can become
separable in the finite time due to the joint action of collective
dephasing and the external driving fields.

Meanwhile, there have been several proposals for controlling the
entanglement of the qubits in the presence of dephasing or
dissipation, such as quantum error-correcting approach
\cite{Shor1995,Steane1996,Knill1997}, quantum error-avoiding
approach \cite{Duan1997,Lidare1998}, and loop control strategies
\cite{Wiseman1993} etc. Here, we present a possible way to
preserve the entanglement of two qubits initially in the fragile
entangled state under the collective dephasing environment. It is
shown that the external local driving field with an appropriate
finite action time can effectively transform the fragile entangled
state into a robust entangled state.

The standard two-qubit Werner state is defined by
\cite{Werner1989} \be
\rho_W=r|\Psi^{-}\rangle\langle\Psi^{-}|+\frac{1-r}{4}I\otimes{I},
\ee where $r\in[0,1]$, and $|\Psi^{-}\rangle$ is the singlet state
of two qubits. $I$ is the identity operator of a single qubit.
Recently, definition (1) is generalized to include the following
states of two qubits \cite{Munro2001,Ghosh2001,Wei2003} \be
\rho^{'}_W=r|M\rangle\langle{M}|+\frac{1-r}{4}I\otimes{I}, \ee
where $|M\rangle$ are any two-qubit maximally entangled states.
Both the Werner state (1) and the Werner-like state (2) are very
important in quantum information. The Werner state (1) is highly
symmetric and $SU(2)\otimes{S}U(2)$ invariant
\cite{Bennett1996,Werner1989}.

The collective dephasing can be described by the master equation
\cite{Lidare1998} \be
\frac{\partial\hat{\rho}}{\partial{t}}=\frac{\gamma}{2}(2\hat{J}_{z}\hat{\rho}\hat{J}_{z}-\hat{J}^2_{z}\hat{\rho}-\hat{\rho}\hat{J}^2_{z}),
\ee where $\gamma$ is the dephasing rate. $\hat{J}_{z}$ are the
collective spin operator defined by \be
\hat{J}_{z}=\sum^{2}_{i=1}\hat{\sigma}^{(i)}_{z}/2,\ee where
$\hat{\sigma}_z$ for each qubit is defined by
$\hat{\sigma}_{z}=|1\rangle\langle{1}|-|0\rangle\langle{0}|$.
Firstly, it is obvious that the standard Werner state (1) is
completely decoupled from the collective dephasing. So in the
presence of collective dephasing, the state (1) belongs to the
decoherence-free subspace, and it maintain its entanglement
invariant. Then, we want to know how the collective dephasing
affects the Werner-like states defined by Eq.(2). For simplicity,
we only consider three states defined by Eq.(2) in which the
maximally entangled states are the Bell triglet states. If two
qubits are initially in the Werner-like state \be
\rho^{'}_{W}=r|\Psi^{+}\rangle\langle\Psi^{+}|+\frac{1-r}{4}I\otimes{I},
\ee where $|\Psi^{+}\rangle$ is one of the Bell states
$|\Psi^{+}\rangle=\frac{\sqrt{2}}{2}(|10\rangle+|01\rangle)$, the
collective dephasing does not change the form of
$r|\Psi^{+}\rangle\langle\Psi^{+}|+\frac{1-r}{4}I\otimes{I}$. So
both the standard Werner state and the Werner-like state
$r|\Psi^{+}\rangle\langle\Psi^{+}|+\frac{1-r}{4}I\otimes{I}$
belong to the decoherence-free subspace of the collective
dephasing. They are robust states against the collective
dephasing. Are all of the Werner-like states robust states against
the collective dephasing? The answer is no. Now we consider
another Werner-like state \be
\rho^{\pm}_{W}=r|\Phi^{\pm}\rangle\langle\Phi^{\pm}|+\frac{1-r}{4}I\otimes{I},
\ee where $|\Phi^{\pm}\rangle$ is the Bell states
$|\Phi^{\pm}\rangle=\frac{\sqrt{2}}{2}(|11\rangle\pm|00\rangle)$.
It is assumed that the initial state of master equation (3) is
$\rho^{\pm}_{W}$. Then its time evolution density matrix can be
expressed as \beqa
\rho^{\pm}_{W}(t)=re^{-2\gamma{t}}|\Phi^{\pm}\rangle\langle\Phi^{\pm}|+\frac{1-r}{4}I\otimes{I}\nonumber\\
+\frac{r}{2}(1-e^{-2\gamma{t}})|11\rangle\langle11|+\frac{r}{2}(1-e^{-2\gamma{t}})|00\rangle\langle00|.
\eeqa In order to quantify the degree of entanglement, we adopt
the concurrence $C$ defined by Wooters \cite{Woo1998}. The
concurrence varies from $C=0$ for an unentangled state to $C=1$
for a maximally entangled state. For two qubits, in the "Standard"
eigenbasis: $|1\rangle\equiv|11\rangle$,
$|2\rangle\equiv|10\rangle$, $|3\rangle\equiv|01\rangle$,
$|4\rangle\equiv|00\rangle$, the concurrence may be calculated
explicitly from the following: \be
C=\max\{\lambda_1-\lambda_2-\lambda_3-\lambda_4,0\}, \ee where the
$\lambda_{i}$($i=1,2,3,4$) are the square roots of the eigenvalues
\textit{in decreasing order of magnitude} of the "spin-flipped"
density matrix operator
$R=\rho_s(\sigma^{y}\otimes\sigma^{y})\rho^{\ast}_s(\sigma^{y}\otimes\sigma^{y})$,
where the asterisk indicates complex conjugation. The concurrence
related to the density matrix $\rho^{\pm}_{W}(t)$ can be written
as \be C(t)=\max(0,\frac{r-1}{2}+re^{-2\gamma{t}}). \ee From
Eq.(9), we can know that, different from the standard Werner state
described by Eq.(1) and
$r|\Psi^{+}\rangle\langle\Psi^{+}|+\frac{1-r}{4}I\otimes{I}$, the
Werner-like state
$r|\Phi^{\pm}\rangle\langle\Phi^{\pm}|+\frac{1-r}{4}I\otimes{I}$
rapidly loses its entanglement in the presence of collective
dephasing. The threshold time $t_c$ beyond which the entanglement
of two qubits completely disappears can be obtained as \be
t_c=-\frac{1}{2\gamma}\ln[\frac{1-r}{2r}]. \ee Being similar to
the results in Ref.\cite{Yu2004}, the Werner-like state described
by Eq.(5) is completely disentangled in a finite time due to the
collective dephasing if the initial parameter $r\neq1$. It is also
interesting to investigate how the collective dephasing affects
the mixedness defined by $M=\frac{4}{3}(1-{\mathrm{Tr}}\rho^2)$.
The mixedness of the time evolution density matrix (6) can be
calculated as \be
M(t)=1-\frac{r^2}{3}-\frac{2r^2}{3}e^{-4\gamma{t}}. \ee When
$\gamma{t}\rightarrow\infty$, the final mixedness of the state in
Eq.(6) equals to $1-\frac{r^2}{3}$. It is shown that the state
$r|\Phi^{\pm}\rangle\langle\Phi^{\pm}|+\frac{1-r}{4}I\otimes{I}$
loses its purity in the collective dephasing. The larger the
parameter $r$, the smaller the final mixedness.

In the following, we attempt to discuss how the collective
dephasing affects the Bell violation of the Werner-like states.
Bell violation is not an entanglement measure. Those states
violating the Bell inequality must be nonseparable. However,
certain kinds of entangled states may not violate the Bell
inequality. The most commonly discussed Bell inequality is the
CHSH inequality \cite{Bell1965,CHSH}. The CHSH operator reads \be
\hat{B}=\vec{a}\cdot\vec{\sigma}\otimes(\vec{b}+\vec{b^{\prime}})\cdot\vec{\sigma}
+\vec{a^{\prime}}\cdot\vec{\sigma}\otimes(\vec{b}-\vec{b^{\prime}})\cdot\vec{\sigma},
\ee where $\vec{a},\vec{a^{\prime}},\vec{b},\vec{b^{\prime}}$ are
unit vectors. In the above notation, the Bell inequality reads \be
|\langle\hat{B}\rangle|\leq2. \ee The maximal amount of Bell
violation of a state $\rho$ is given by \cite{Horo1995} \be
{\mathcal{B}}=2\sqrt{\lambda+\tilde{\lambda}}, \ee where $\lambda$
and $\tilde{\lambda}$ are two largest eigenvalues of
$T^{\dagger}_{\rho}T_{\rho}$. The matrix $T_{\rho}$ is determined
completely by the correlation functions being a $3\times3$ matrix
whose elements are
$(T_{\rho})_{nm}={\mathrm{Tr}}(\rho\sigma_{n}\otimes\sigma_{m})$.
Here, $\sigma_1\equiv\sigma_x$, $\sigma_2\equiv\sigma_y$, and
$\sigma_3\equiv\sigma_z$ denote the usual Pauli matrices. We call
the quantity $\mathcal{B}$ the maximal violation measure, which
indicates the Bell violation when ${\mathcal{B}}>2$ and the
maximal violation when ${\mathcal{B}}=2\sqrt{2}$. For the density
matrix $\rho^{\pm}_{W}(t)$ in Eq.(6), $\lambda+\tilde{\lambda}$
can be written as follows \be
\lambda+\tilde{\lambda}=r^2(1+e^{-4\gamma{t}}). \ee The threshold
time $t^{b}_c$ beyond which the collective dephasing Werner-like
state does not violate the Bell-CHSH inequality can be given by
\be t^{b}_c=-\frac{1}{4\gamma}\ln(\frac{1-r^2}{r^2}).\ee
Eqs.(14-15) show that, if $r\neq1$, i.e. the initial state is not
a maximally entangled state, the dephasing Werner-like state in
Eq.(7) rapidly loses its nonlocality in a finite time. The smaller
the initial entanglement, the more rapidly the nonlocality
completely disappears. In Fig.1, the threshold time $t_c$ and
$t^{b}_c$ concerning the complete disentanglement and the
disappearance of nonlocality of the state (7) respectively, have
been plotted as the function of the parameter $r$. It implies that
both the entanglement and nonlocality are completely destroyed in
a finite time if the initial state is not pure. While the
collective dephasing can not completely destroy nonlocality and
entanglement of the pure Bell states $|\Phi^{\pm}\rangle$ in the
finite time. According to two threshold times, we can classify the
Werner-like state (6) by making use of the parameter $r$ or the
initial mixedness. If $r=0$, the state (6) reduces to the
maximally mixed state. In the range of $0<r\leq\frac{1}{3}$, the
states in Eq.(6) are separable. In the range of
$\frac{1}{3}<r\leq\frac{\sqrt{2}}{2}$, they are entangled but not
nonlocal. In the range of $\frac{\sqrt{2}}{2}<r<1$, they are
inseparable and nonlocal, and both their nonlocality and
entanglement can be completely destroyed by the collective
dephasing. Recently, Yu and Eberly have shown a novel phenomenon
that, under the influence of pure vacuum noise two entangled
qubits become completely disentangled in a finite-time
\cite{Yu2004}. In a specific example they have found the time to
be given by $\ln(1+\frac{\sqrt{2}}{2})$ times the usual
spontaneous lifetime. Here, we have obtained a similar result that
in the presence of collective dephasing, two initial mixed
entangled qubits become completely disentangled in a finite-time.
It is conjectured that, in this case, the initial mixedness is an
essential fact whether two entangled qubits become completely
disentangled in a finite-time or not. In fact, previous work
concerning the interaction between a initial mixed qubit and the
thermal field in the presence of phase decoherence have revealed
some analogical results \cite{Li2003}.

\begin{figure}
\centerline{\includegraphics[width=2.5in]{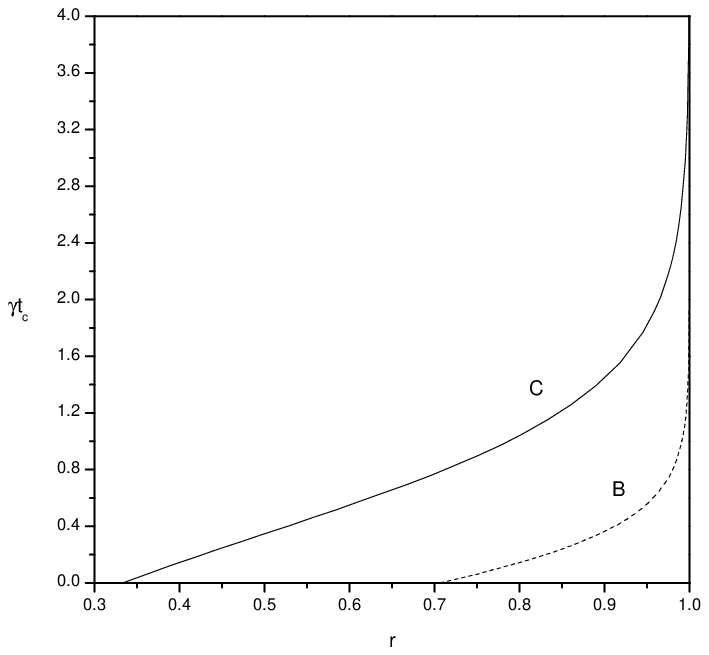}}
\caption{ The threshold time $\gamma{t}_c$ and $\gamma{t}^{b}_c$
of the concurrence and the maximal Bell violation respectively,
are plotted as the function of the parameter $r$ for the
collective dephasing Werner-like states in Eq.(7). (Solid line)
$\gamma{t}_c$ in Eq.(10); (Dash line) $\gamma{t}^{b}_c$ in
Eq.(16). It is shown that both $\gamma{t}_c$ and $\gamma{t}^{b}_c$
increase with $r$.}
\end{figure}

It is also very interesting to investigate how an external driving
field affects the complete disentanglement time in this situation.
If the external driving fields are taken into account, the master
equation (3) should be replaced by \be
\frac{\partial\hat{\rho}}{\partial{t}}=-\frac{i}{2}[\Omega_1\hat{\sigma}^{(1)}_x+\Omega_2\hat{\sigma}^{(2)}_x,\hat{\rho}]+\frac{\gamma}{2}(2\hat{J}_{z}\hat{\rho}\hat{J}_{z}-\hat{J}^2_{z}\hat{\rho}-\hat{\rho}\hat{J}^2_{z}),
\ee where $\Omega_{i}$ ($i=1,2$) are the intensity of the external
driving field acted on the $i$th qubit. $\hat{\sigma}_x$ for each
qubit are defined by
$\hat{\sigma}_{x}=|1\rangle\langle{0}|+|0\rangle\langle{1}|$. In
the case with $\Omega_1=\Omega_2=\Omega$, the standard Werner
state is still decoupled from the above master equation, while
other Werner-like states lose their entanglement. In order to know
whether
$r|\Psi^{+}\rangle\langle\Psi^{+}|+\frac{1-r}{4}I\otimes{I}$ and
$r|\Phi^{\pm}\rangle\langle\Phi^{\pm}|+\frac{1-r}{4}I\otimes{I}$
can be completely disentangled in a finite time or not, it is
sufficient to verify Bell states $|\Psi^{+}\rangle$ and
$|\Phi^{\pm}\rangle$ are completely disentangled in their
evolutions governed by Eq.(17). In Fig.2, the dynamical behaviors
of concurrence of two qubits initially in Bell triplet states
governed by Eq.(17) are displayed. It is shown that all of Bell
triplet states are completely disentangled in a finite time, which
imply all of the Werner-like states defined by Eq.(5) and Eq.(7)
are finite-time disentangled in the joint action of the collective
dephasing and external driving fields. From Fig.2, it can be
observed that $|\Phi^{-}\rangle$ most rapidly loses its
entanglement among all of Bell triplet states.
\begin{figure}
\centerline{\includegraphics[width=2.5in]{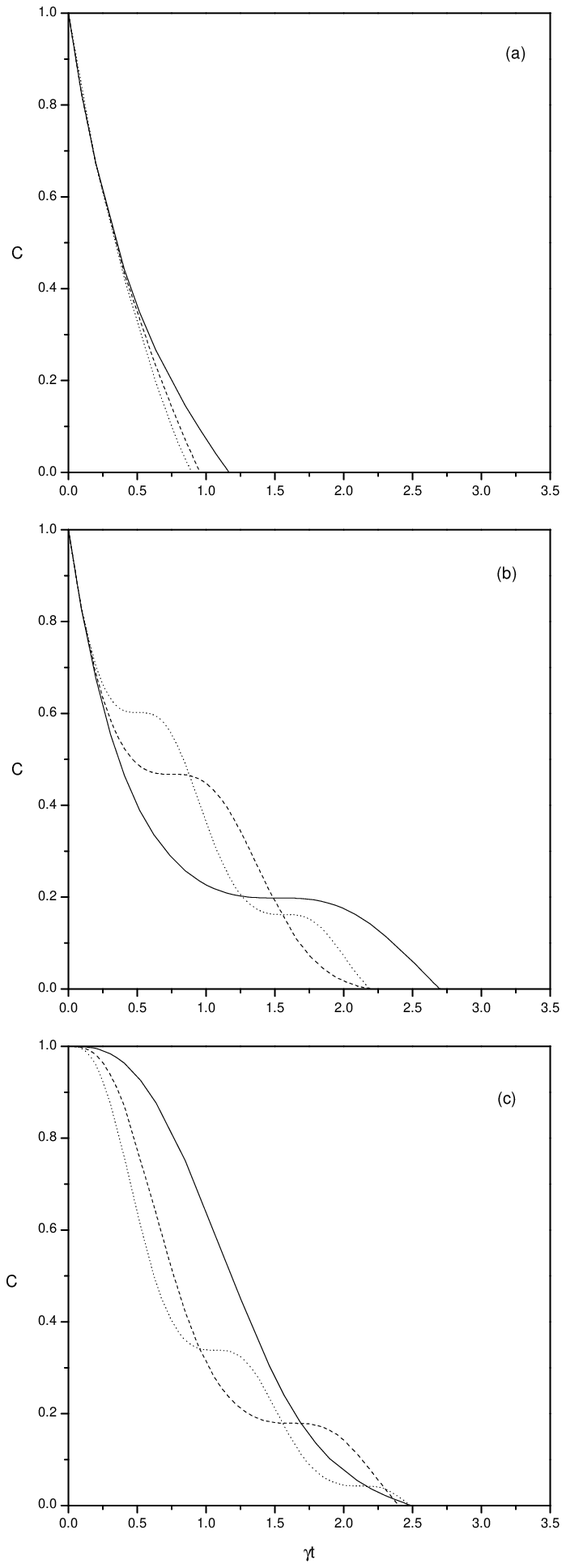}}
\caption{The dynamical behaviors of concurrence of two qubits
initially in Bell triplet states governed by Eq.(17) are displayed
as the function of the dephasing time $\gamma{t}$ for different
values of $\Omega_1=\Omega_2=\Omega$. (Solid line)
$\Omega=\gamma$, (Dash line) $\Omega=2\gamma$, (Dot line)
$\Omega=3\gamma$. (a) two qubits are initially in
$|\Phi^{-}\rangle$; (b) two qubits are initially in
$|\Phi^{+}\rangle$; (c) two qubits are initially in
$|\Psi^{+}\rangle$. It is shown that all of Bell triplet states
become separable in different finite times in the joint action of
collective dephasing and driving fields. In the case without the
driving fields, Bell triplet states can not become separable in
finite time.}
\end{figure}
In the case with $\Omega_1\neq\Omega_2$, the Bell singlet state is
no longer decoupled from the Eq.(17). In Fig.3, we show that four
Bell states become complete disentanglement in the finite time in
the situation with $\Omega_1=\gamma$ and $\Omega_2=0$. The
partially driving field acting on the qubit 1 destroys the
symmetry of the pure collective dephasing and forces the Bell
singlet state out of the decoherence-free subspace. Nevertheless,
$|\Psi^{\pm}\rangle$ are more robust than $|\Phi^{\pm}\rangle$ in
this case, and the threshold time corresponding to the complete
disentanglement of two qubits initially in $|\Psi^{\pm}\rangle$ is
about twice as large as the one corresponding to the complete
disentanglement of two qubits initially in $|\Phi^{\pm}\rangle$.
\begin{figure}
\centerline{\includegraphics[width=2.5in]{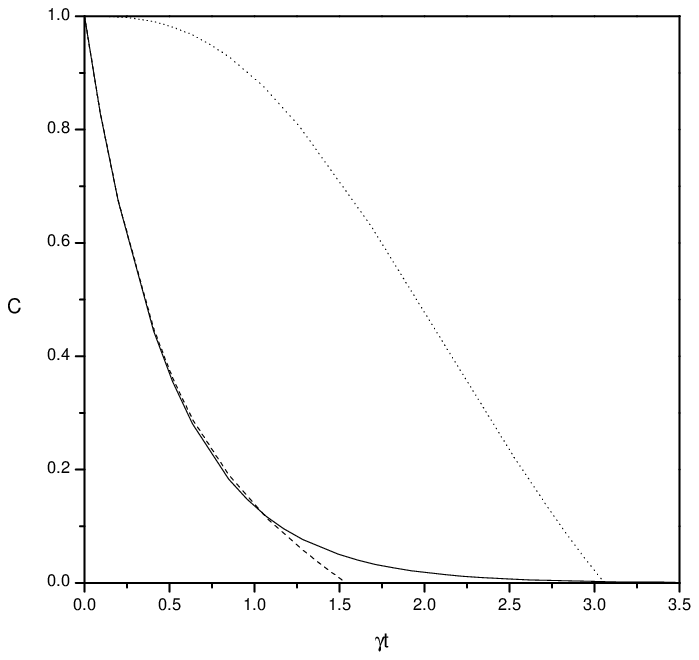}}
\caption{The dynamical behaviors of concurrence of two qubits
initially in four Bell states governed by the master equation (17)
are displayed as the function of the dephasing time $\gamma{t}$
with $\Omega_1=\gamma$ and $\Omega_2=0$. (Dash line) two qubits
are initially in $|\Phi^{\pm}\rangle$; (Dot line) two qubits are
initially in $|\Psi^{\pm}\rangle$; (Solid line) for comparison, we
display the asymptotic disentanglement of two qubits initially in
$|\Phi^{\pm}\rangle$ in the presence of the pure collective
dephasing without any external driving fields.}
\end{figure}

In what follows, we briefly discuss a possible scheme to prevent
the fragile entangled states from complete disentanglement under
the action of collective dephasing. We assume that only one of two
qubits is driven by a time-dependent external field. For
simplicity, the time dependence of the external driving field is
suggested to be the form of the unit step function. The dynamics
of two qubits can be described by the following master equation
\be
\frac{\partial\hat{\rho}}{\partial{t}}=-\frac{i}{2}[\zeta_1(t)\hat{\sigma}^{(1)}_x,\hat{\rho}]+\frac{\gamma}{2}(2\hat{J}_{z}\hat{\rho}\hat{J}_{z}-\hat{J}^2_{z}\hat{\rho}-\hat{\rho}\hat{J}^2_{z}),
\ee where $\zeta_{1}(t)=\zeta_1\Theta(T-t)$ is the intensity of
the time-dependent external driving field acted on the qubit 1,
and $\Theta(x)$ is the unit step function and equals one for
$x\geq0$ and equals zero for $x<0$. In previous experimental
verification of the decoherence-free subspace \cite{Kwiat2000},
the external driving field and the collective dephasing are not
simultaneously acted on the qubits. It is obvious that the local
external driving field $\zeta_1(t)\hat{\sigma}^{(1)}_x$ can
interconvert the Bell states without the simultaneous presence of
collective dephasing. Nevertheless, it is desirable to determine
the influence of the variation of $T$ on the eventual stationary
state entanglement when the driving field and the collective
dephasing are simultaneously acted on the qubits in some realistic
situations. Our numerical calculations show that the external
field with an appropriate value of $T$ can transform the initial
fragile entangled state $|\Phi^{+}\rangle$ into a stationary
entangled state even if the collective dephasing is always
presence. In Fig.4, the stationary state concurrence $C_s$ and the
stationary state Bell violation $|B^{(s)}|_{max}$ of two qubits
are plotted as the function of the parameter $\gamma{T}$. It is
shown that, if only the value of $\zeta_1/\gamma$ is large enough,
one can maintain the entanglement and Bell violation of two qubits
initially in a very fragile entangled state by making use of the
local driving field with an appropriate action time $T$. We can
see that the stationary state entanglement firstly increases with
$\gamma{T}$, and achieves a local maximal value, then decreases
with $\gamma{T}$. Similar behaviors are repeated again when the
scaled action time $\gamma{T}$ is further enlarged. The stationary
state Bell violation also oscillates with $\gamma{T}$. The above
calculations show one can effectively transform the fragile
entangled state into a robust entangled state. This is meaningful
and very important in many areas of quantum information processes.

\begin{figure}
\centerline{\includegraphics[width=2.5in]{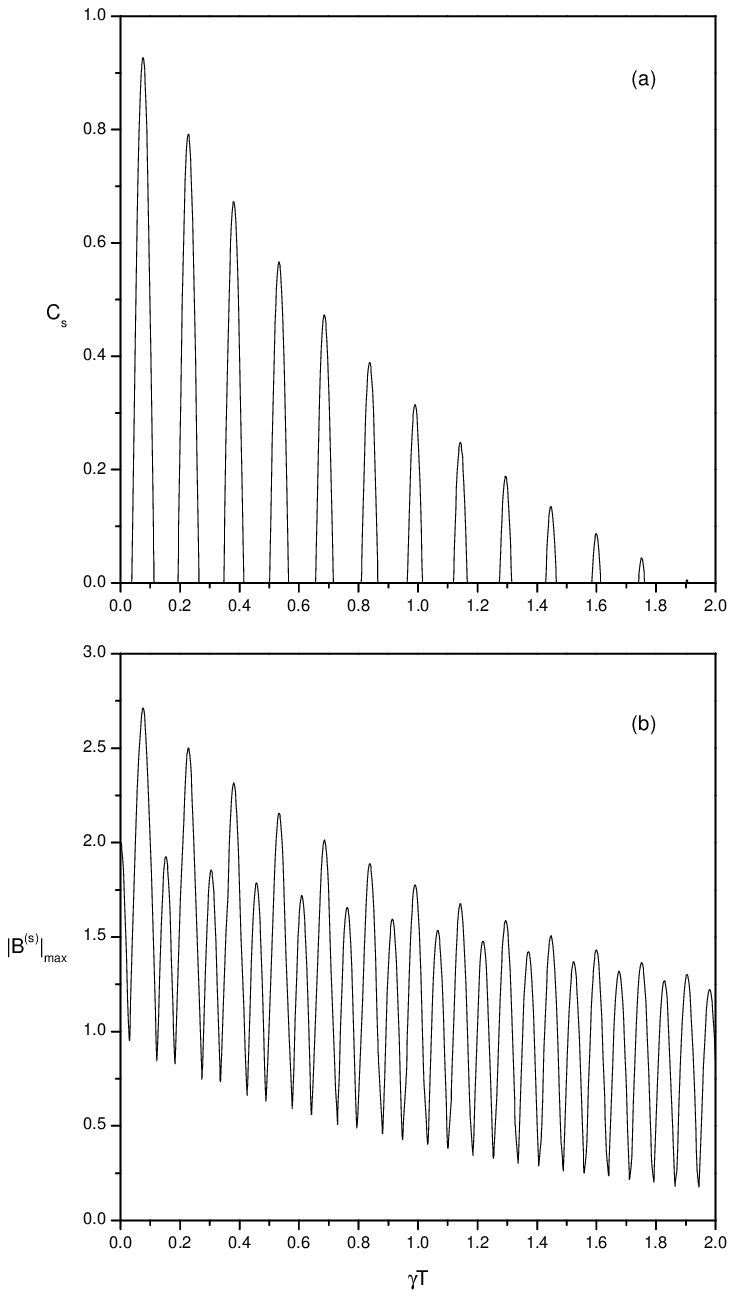}}
\caption{The stationary state concurrence $C_s$ (a) and the
stationary state Bell violation $|B^{(s)}|_{max}$ (b) of two
qubits initially in the Bell states $|\Phi^{+}\rangle$ governed by
the master equation (17) are displayed as the function of the
scaled action time $\gamma{T}$ with $\zeta_1/\gamma=41.25$. In
this case, when $\gamma{T}>2$, there is not any stationary state
entanglement between two qubits.}
\end{figure}

Finally, we attempt to discuss the theoretical results about the
fragile and robust Werner state in collective dephasing by
comparing them with previous experimental results. In
ref.\cite{Kwiat2000}, Kwiat et al. have presented the experimental
verification of decoherence-free subspace in the system of
polarized photons. For investigating the influence of collective
dephasing on the initial state, one can adopt the general fidelity
$F(\rho_i,\rho_f)\equiv[{\mathrm{Tr}}(\sqrt{\rho_i}\rho_f\sqrt{\rho_i})^{1/2}]^2$
\cite{Jozsa1994}, where $\rho_i$ is the initial state and $\rho_f$
the final state. The fidelity between the initial Werner-like
state $\rho^{\pm}_W$ in Eq.(6) and the corresponding dephasing
state $\rho^{\pm}_W(t)$ in Eq.(7) can be calculated as \beqa
F_W&=&\frac{1}{16}[2(1-r)+\sqrt{(1+3r)(1+r+2re^{-2\gamma{t}})}\nonumber\\
&&+\sqrt{(1-r)(1+r(1-2e^{-2\gamma{t}}))}]^2, \eeqa which decreases
with $r$ and $t$. In this case, the minimal value of $F_W$ is 0.5
at $r=1$ and $t\rightarrow\infty$. The collective dephasing in the
experiment of Kwiat et al. can be described by the following
master equation which is equivalent to the Eq.(3) under the local
unitary transformation. \be
\frac{\partial\hat{\rho}}{\partial{t}}=\frac{\gamma}{2}(2\hat{J}_{\theta}\hat{\rho}\hat{J}_{\theta}-\hat{J}^2_{\theta}\hat{\rho}-\hat{\rho}\hat{J}^2_{\theta}),
\ee where $\hat{J}_{\theta}$ are the collective spin operator
defined by \be
\hat{J}_{\theta}=\sum^{2}_{i=1}\hat{\sigma}^{(i)}_{\theta}/2,\ee
where $\hat{\sigma}_{\theta}$ for each qubit is defined by
$\hat{\sigma}_{\theta}=\cos2\theta\hat{\sigma}_z+\sin2\theta\hat{\sigma}_x$.
Throughout the following calculations, we choose
$\theta=17^{\circ}$ for representing the realistic situation in
Ref.\cite{Kwiat2000}. The fidelity between the initial Werner-like
states and corresponding stationary state of Eq.(20) for three
different initial states has been calculated and the results are
depicted in Fig.5. It was found that the fidelities decrease with
$r$ and the theoretical results excellently agree with the
experimental data in Ref.\cite{Kwiat2000}. By comparing two
table.1 in present paper and in Ref.\cite{Kwiat2000}, it may be
interesting that the initial states in \cite{Kwiat2000} look like
the Werner or Werner-like states with very high purity. In Fig.6,
the evolution of fidelities are investigated for three different
initial states. It is shown that the fidelity decreases with
dephasing time and eventually stays at a fixed value, which
implies that these chosen initial states are out of the
decoherence-free subspace of Eq.(20). It is shown that evolving
fidelity of the initial state
$0.99|\Phi^{-}\rangle\langle\Phi^{-}|+\frac{1}{400}I\otimes{I}$ is
larger than the one of
$0.99|\Phi^{+}\rangle\langle\Phi^{+}|+\frac{1}{400}I\otimes{I}$ in
short time.

\begin{figure}
\centerline{\includegraphics[width=2.5in]{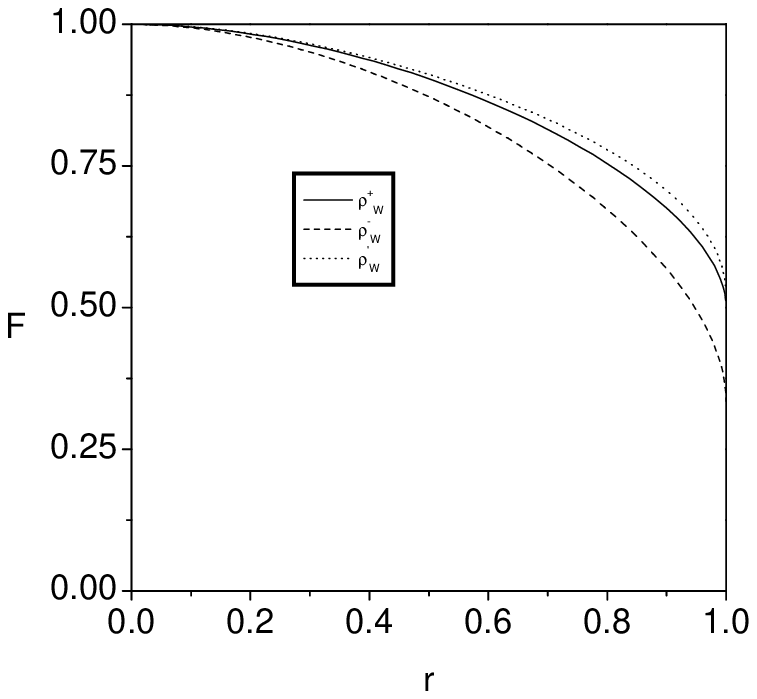}}
\caption{The fidelity
$F\equiv[{\mathrm{Tr}}(\sqrt{\rho_i}\rho_f\sqrt{\rho_i})^{1/2}]^2$
for three kinds of initial states and their corresponding complete
dephasing states are plotted as the function of the parameter $r$.
(Solid line)
$\rho_i=\rho^{+}_W=r|\Phi^{+}\rangle\langle\Phi^{+}|+\frac{1-r}{4}I\otimes{I}$;
(Dash line)
$\rho_i=\rho^{-}_W=r|\Phi^{-}\rangle\langle\Phi^{-}|+\frac{1-r}{4}I\otimes{I}$;
(Dot line)
$\rho_i=\rho^{\prime}_W=r|\Psi^{+}\rangle\langle\Psi^{+}|+\frac{1-r}{4}I\otimes{I}$.
It is shown that $F(\rho^{\prime}_W)>F(\rho^{+}_W)>F(\rho^{-}_W)$
for any non-zero values of $r$, where we have simplified the
expression $F(\rho_i,\rho_f)$ as $F(\rho_i)$. This simplification
has also been adopted in table.1. }
\end{figure}

\begin{table}
\centerline{\includegraphics[width=2.5in]{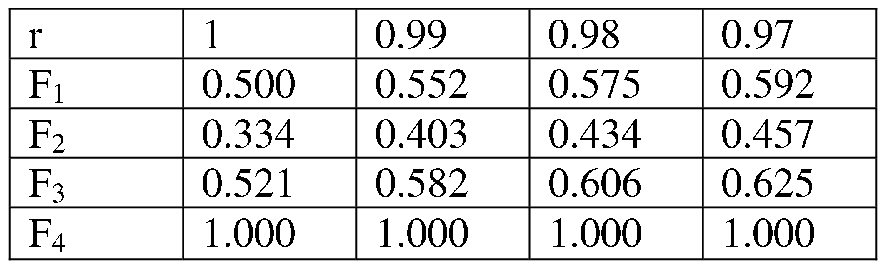}}
\caption{The fidelities $F_i$ ($i=1,2,3,4$) of the initial states
$\rho^{+}_W$ (in Eq.(6)), $\rho^{-}_W$ (in Eq.(6)),
$\rho^{\prime}_W$ (in Eq.(5)), $\rho_W$ (in Eq.(1)) and their
corresponding stationary states of Eq.(20) in order are listed for
different values of $r$. $F_1\equiv{F(\rho^{+}_W)}$;
$F_2\equiv{F(\rho^{-}_W)}$; $F_3\equiv{F(\rho^{\prime}_W)}$;
$F_4\equiv{F(\rho_W)}$. We can see that present theoretical
calculations based on the Eq.(20) and the initial Werner-like
state assumption well agree with the experimental data in table.1
of Ref.\cite{Kwiat2000}.}
\end{table}

\begin{figure}
\centerline{\includegraphics[width=2.5in]{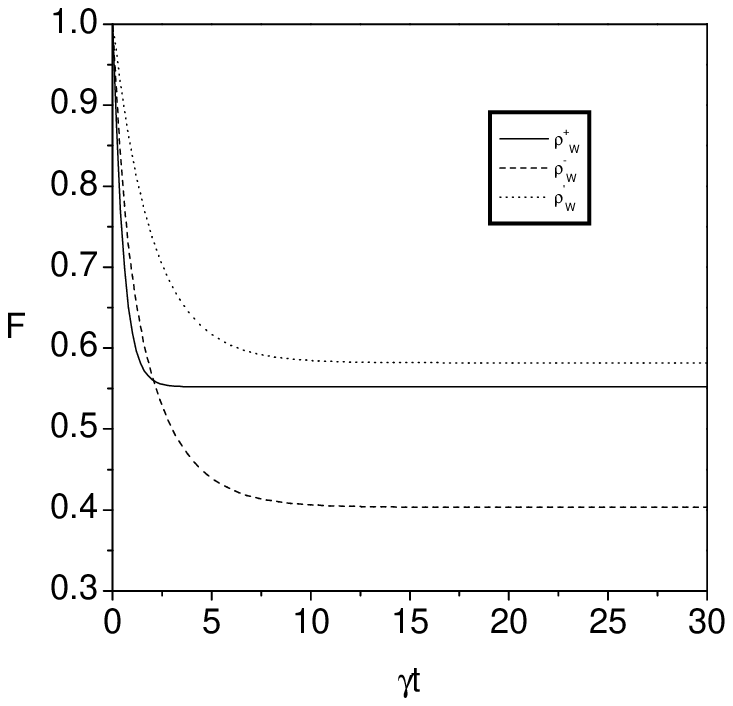}}
\caption{The fidelities for three kinds of initial states and
their corresponding evolving states governed by Eq.(20) are
plotted as the function of $\gamma{t}$. The initial Werner-like
states with $r=0.99$ are chosen for calculating all of three
curves.}
\end{figure}

In summary, we investigate the concurrence and Bell violation of
the standard Werner state or Werner-like states in the presence of
collective dephasing. By making use of the analytical expressions
of the concurrence and Bell violation obtained in the present
paper, we find that the standard Werner state and certain kinds of
Werner-like states are robust against the collective dephasing,
and some kinds of Werner-like states is fragile and becomes
completely disentangled in a finite-time. The threshold time of
complete disentanglement of the fragile Werner-like states is
analyzed. We conjecture that the initial mixedness is an important
fact to determine whether two entangled qubits become completely
disentangled in a finite-time or not in the pure collective
dephasing. Moreover, the threshold time concerning the complete
disappearance of the Bell violation of some kinds of fragile
Werner-like states is also obtained. Furthermore, we investigate
how an external driving field affects the completely
disentanglement time and clarify that the pure maximally entangled
states of two qubits can become separable in the finite time in
this situation. Since the standard Werner state or Werner-like
states play a special role in some quantum information processes,
such as quantum teleportation, our results may have potential
applications in quantum teleportation
\cite{Bennett1993,Bouwmester1997} or other remote quantum
information processes.

We also discuss the possible way to transform the fragile
entangled state into the robust entangled state in the collective
dephasing environment. It is shown that a local external driving
field with an appropriate finite action time can effectively
maintain both the entanglement and Bell violation of two qubits
even if the collective dephasing is presence from beginning to
end.

Recently, the quantum information processes in the presence of the
collective dephasing have intrigued much attention
\cite{Khodjasteh2002,Yu2002,Yu2003,Hill2004,Ollerenshaw2003}.
Khodjasteh and Lidar have investigated the universal
fault-tolerant quantum computation in the presence of spontaneous
emission and collective dephasing \cite{Khodjasteh2002}. Hill and
Goan have studied the effect of dephasing on proposed quantum
gates for the solid-state Kane quantum computing architecture
\cite{Hill2004}. In the future work, it may be very interesting to
apply the present results to discuss the influence of collective
dephasing on gate fidelity of remote quantum computation.

Finally, we also compare the theoretical results about the
fidelity of the initial Werner-like state in the presence of
collective dephasing with recent experimental data of Kwiat et al
\cite{Kwiat2000}. It is shown that they are consistent with each
other.

\section*{ACKNOWLEDGMENT}
This project was supported by the National Natural Science
Foundation of China (Project NO. 10174066).

\bibliographystyle{apsrev}
\bibliography{refmich,refjono}

\end{document}